\documentclass[floatfix,twocolumn,showpacs,preprintnumbers,amsmath,amssymb,prl]{revtex4}

\usepackage{graphicx}
\usepackage{epstopdf}
\usepackage{dcolumn}
\usepackage{rotating}
\usepackage{setspace} 
\usepackage{hyperref}

\begin{document}

\title{Emergent Kinetics and Fractionalized Charge in 1D Spin-Orbit Coupled Flatband Optical Lattices}
\author{Fei Lin$^{1}$, Chuanwei Zhang$^{2}$, and V. W. Scarola$^{1}$}
\affiliation{$^{1}$Department of Physics, Virginia Tech, Blacksburg, Virginia 24061 USA}
\affiliation{$^{2}$Department of Physics, The University of Texas at Dallas, Richardson, Texas, 75080 USA}

\date{\today}

\begin{abstract}
Recent ultracold atomic gas experiments implementing synthetic spin-orbit coupling allow access to flatbands that emphasize interactions. We model spin-orbit coupled fermions in a one-dimensional flat band optical lattice. We introduce an effective Luttinger-liquid theory to show that interactions generate collective excitations with emergent kinetics and fractionalized charge, analogous to properties found in the two-dimensional fractional quantum Hall regime. Observation of these excitations would provide an important platform for exploring exotic quantum states derived solely from interactions.  
\end{abstract}


\maketitle

\noindent
{\it Introduction.} -- Emergent quantum states derived from interactions can exhibit rich structure because they are, by definition, not adiabatically connected to the underlying single-particle states.  Two-dimensional (2D) electron gases placed in a strong magnetic field offer seminal examples.  In the absence of a magnetic field, 2D electrons typically demonstrate Fermi-liquid behavior, but a strong magnetic field, the fractional quantum Hall (FQH) limit \cite{tsui:1982}, would seem to prevent the formation of a Fermi liquid.  This regime is defined by an absence of single-particle kinetic energy that leaves inter-particle interactions to generate many-body quantum states in a flatband (the lowest Landau level).  However, it is now well known that interesting properties, such as fractional charge from screening and other kinetic effects \cite{laughlin:1983,goldman:1995,dassarma:1997}, emerge from interactions in the FQH regime.  The remarkable fact that application of an external field first suppresses single-particle properties to leave interactions to generate similar emergent properties leads to a natural question:  Can these emergent mechanisms manifest in other contexts?  Flatbands in one dimension offer a logical analogue \cite{westerberg:1993,seidel:2006,jansen:2008,nakamura:2012}.

The Luttinger-liquid paradigm \cite{haldane:1981,fisher:1997,miranda:2003,gimarchy:2003} captures the physics of many one-dimensional (1D) models.  It predicts excitations with, e.g., fractionalized charge arising from competition between interactions and kinetic energy.  External fields could, in analogy to 2D magnetic fields, be constructed to quench kinetics in 1D, but the absence of kinetics in 1D flatbands would appear to rule out Luttinger-liquid behavior.  

In this Letter, we show that kinetics, fractionalized charge excitations, and other Luttinger-liquid-like properties emerge solely from interactions in experimentally feasible 1D flatband models. Our proposal relies on recent experimental progress \cite{lin:2011,wang:2012,cheuk:2012,zhang:2012,fu:2013,williams:2013} in engineering synthetic spin-orbit coupling (SOC) for ultracold atomic gases \cite{bloch:2008}.  These experiments show that Raman beams can be used to dress atoms with spin-dependent momentums.  Rashba (and/or Dresselhaus) SOCs governing these dressed states \cite{sau:2011,galitski:2013} are tunable to extremes not possible in solids.  Recent work shows that Rashba coupling in a 1D optical lattice \cite{zhang:2013} or gas \cite{zhou:2013,ramachandhran:2013} can be tuned to yield flatbands, a new limit that could play a role analogous to the lowest Landau level \cite{scarola:2013}, but interaction effects in a 1D flat Rashba SOC band remain unexplored.      

We study the impact of interactions between two-component fermions in a flat SOC band in 1D optical lattices.  We find that the SOC elongates single-particle basis states to generate highly non-trivial nearest neighbor (NN) interactions \cite{williams:2012}.  The extended interactions lead to Wigner crystals of spinors with dispersive collective modes.  These excitations are unexpected because they imply kinetics that emerge purely from interactions.  

We predict that these excitations also exhibit fractionalized charge even in the flatband limit.  To show this, we must contend with the fact that the absence of single-particle kinetic energy prevents direct application of the Luttinger-liquid theory.   We find, instead, that emergent kinetics allows us to introduce an effective Luttinger-liquid theory.  We compute the emergent velocities and fractionalized charge of excitations as experimentally verifiable observables. We also estimate the experimental parameters for observing these excitations. Detection of kinetics and fractionalized charge derived solely from interactions in 1D would have important consequences for the study of emergent Luttinger-liquid behavior, in analogy to emergent fractional charge found in the 2D FQH regime.
 
\noindent
{\it Model.} -- We consider an equal population of $N$ two-component fermions in a 1D optical lattice.  We start with a non-interacting Hamiltonian that adds Rashba  SOC to the optical lattice potential \cite{zhang:2013}:
\begin{equation}
\mathcal{H}_{SOC}=\frac{p_x^2}{2m}-sE_R\cos^2(k_L x)+\left(\frac{\hbar k_R}{m}\right ) p_x\sigma_z+\Omega\sigma_x,
\label{spinorbitH}
\end{equation}
where
$p_x$ is the momentum of particles of mass $m$, $s$ is the optical lattice strength, $k_L$ is the optical lattice wave vector, $E_R=\hbar^2k_L^2/2m$ is the recoil energy, $\hbar k_R/m$ is the SOC strength, $\boldsymbol{\sigma}$ are Pauli matrices, and $\Omega$ is the Zeeman field strength.  We work in units of the lattice spacing, $\pi/k_{L}$.

\begin{figure}[t]
\begin{center}
\vspace{-0.2in}
\includegraphics[width=3.8in]{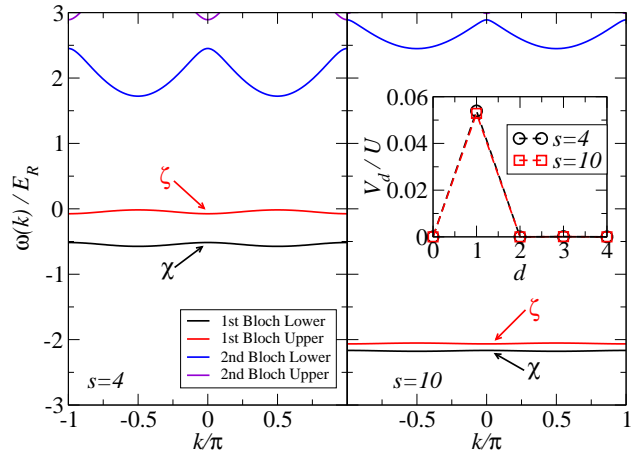}
\vspace{-0.5in}
\end{center}
\caption{(Color online.) (Main panels) Single-particle energy $\omega(k)$ of several lowest Bloch bands due to SOC for $s=4$ (left) and $s=10$ (right).  The ratio of the lowest energy gap to the bandwidth is tuned to $\approx 8$ \cite{supplementarymaterial} in both panels by setting $k_R=k_L/2$, $\Omega=0.22 E_{R}$ for $s=4$, and $\Omega=0.05 E_{R}$ for $s=10$.  (Inset) Diagonal interaction between $\chi$ particles as a function of inter-site distance, $d$, for $s=4E_{R}$ and $s=10E_{R}$ yielding $V_1/U\approx0.0529$.}
\label{Vvsd}
\end{figure}

Figure ~\ref{Vvsd} plots the eigenvalues of $\mathcal{H}_{SOC}$, $\omega(k)$ to show that Eq.~(\ref{spinorbitH}) yields flatbands.  We project into the lowest flatband.  Projection, achieved by considering only $\chi$ particles, is warranted in the presence of an energy gap between the $\chi$ and $\zeta$ bands at low densities \cite{supplementarymaterial}.

We can derive a low-energy Hubbard model of interactions operating in such a flat Rashba band in the tight binding limit.  In the Wannier basis the inter-atom interactions (e.g., $s$-wave contact interactions between alkali atoms) become purely on site.  After projection to the lowest flatband, the on-site Hubbard interaction defines the Hamiltonian of the entire system, and is therefore the focus of our study \cite{supplementarymaterial}:  
 \begin{equation}
H= U\sum_{\{k\}}f_{\{k\}}\chi_{k_1}^{\dagger}
\chi_{k_2}^{\dagger}\chi_{k_3}^{\phantom{\dagger}}\chi_{k_4}^{\phantom{\dagger}},
\label{Hflatband}
\end{equation}
where $U$ is the on-site Hubbard repulsion that defines the only energy scale, $\chi_k^{\dagger}$ creates a fermion at wave vector $k$ in the lowest band, and 
$f_{\{k\}}\equiv L^{-1}\delta_{k_1+k_2, k_3+k_4}\sin(\alpha_{k_1})\cos(\alpha_{k_2})\cos(\alpha_{k_3})\sin(\alpha_{k_4})$.  Here the Kronecker delta implies momentum conservation up to a reciprocal lattice vector and $L$ is the number of lattice sites.

Equation ~(\ref{Hflatband}) is written in terms of lowest flatband-projected particles using a unitary transformation between the original fermions and flatband fermions, so that $\chi$ particles are defined as spinors of the original atoms \cite{supplementarymaterial}.  We define the unitary transformation in terms of optical lattice parameters: $\tan(\alpha_k)=[\omega_1(k)-h(k)]/\Omega$ with $h(k)\equiv-2t\cos(k+k_R)$, and 
$\omega_1(k)\equiv-2t\cos k\cos k_R-\sqrt{4t^2\sin^2k\sin^2k_R+\Omega^2}$. Here $t$ is the NN hopping \cite{supplementarymaterial}.  

%%%%%% VD vs D %%%%%%%%%%%%%%%%
%%%%%% %%%%%%%%%%%%%%%%

Projection to $\chi$ particles generates non-trivial delocalized single-particle basis states.  To see this we Fourier transform $\chi_k$ to real space.  The on-site interaction between the original atoms becomes a longer range interaction between $\chi$ particles. The leading diagonal interaction, $V_d\tilde{n}_i\tilde{n}_{i+d}$, is between NN. Here $\tilde{n}_i^{\vphantom{\dagger}}\equiv\chi_i^{\dagger}\chi_i^{\vphantom{\dagger}}$.  

The inset of Fig.~\ref{Vvsd} shows the interaction strength, $V_d$, between $\chi$ particles.  Different optical lattice depths lead to the same interaction, where $V_d$ falls off quickly with the dominant interaction given by $V_1$, provided the band remains flat \cite{supplementarymaterial}.  The inset shows two key results: (1) The interaction is longer range, and (2) the form of the interaction is robust over a wide range of $s$.  In the following we can therefore focus on $s=10$ without loss of generality.

Equation ~(\ref{Hflatband}) contains a large number of terms, but by considering a few of the largest terms (with strengths $V_{1}$, $t_1^{*}$, and $t_2^{*}$), we argue for intriguing low-energy states.  Leading off-diagonal terms in Eq.~(\ref{Hflatband}) are given by conditional next nearest neighbor (NNN) 
hoppings of $\chi$ particles, i.e., $-|t_1^{*}|\chi_{i+2}^{\dagger}\tilde{n}_{i+1}^{\vphantom{\dagger}}\chi_i^{\vphantom{\dagger}}$ and $|t_2^{*}|\tilde{n}_{i}^{\vphantom{\dagger}}\chi_{i+3}^{\dagger}\chi_{i+1}^{\vphantom{\dagger}}$, where $|t_1^{*}|/U=0.0257$ 
and $|t_2^{*}|/U=0.0015$.  We note that conditional hopping originates entirely from interactions.  The $V_{1}$ term is the strongest and should generate crystal states of spinor $\chi$ particles, but  conditional hoppings can give rise to emergent kinetics in excitations.  We verify this picture below by combining diagonalization with an effective model.

\noindent
{\it Numerical results.} -- To more rigorously study Eq.~(\ref{Hflatband}), we use numerics to explore the low-energy Hilbert space and confine our study to half filling, $N/L$=1/2.  
We note that the absence of kinetic energy excludes the direct use of Luttinger-liquid theory.

%%%%%%%%%%%%% Dispersion %%%%%%%%%%
\begin{figure}[t]
\begin{center}
\vspace{-0.3in}
\includegraphics[width=3.5in]{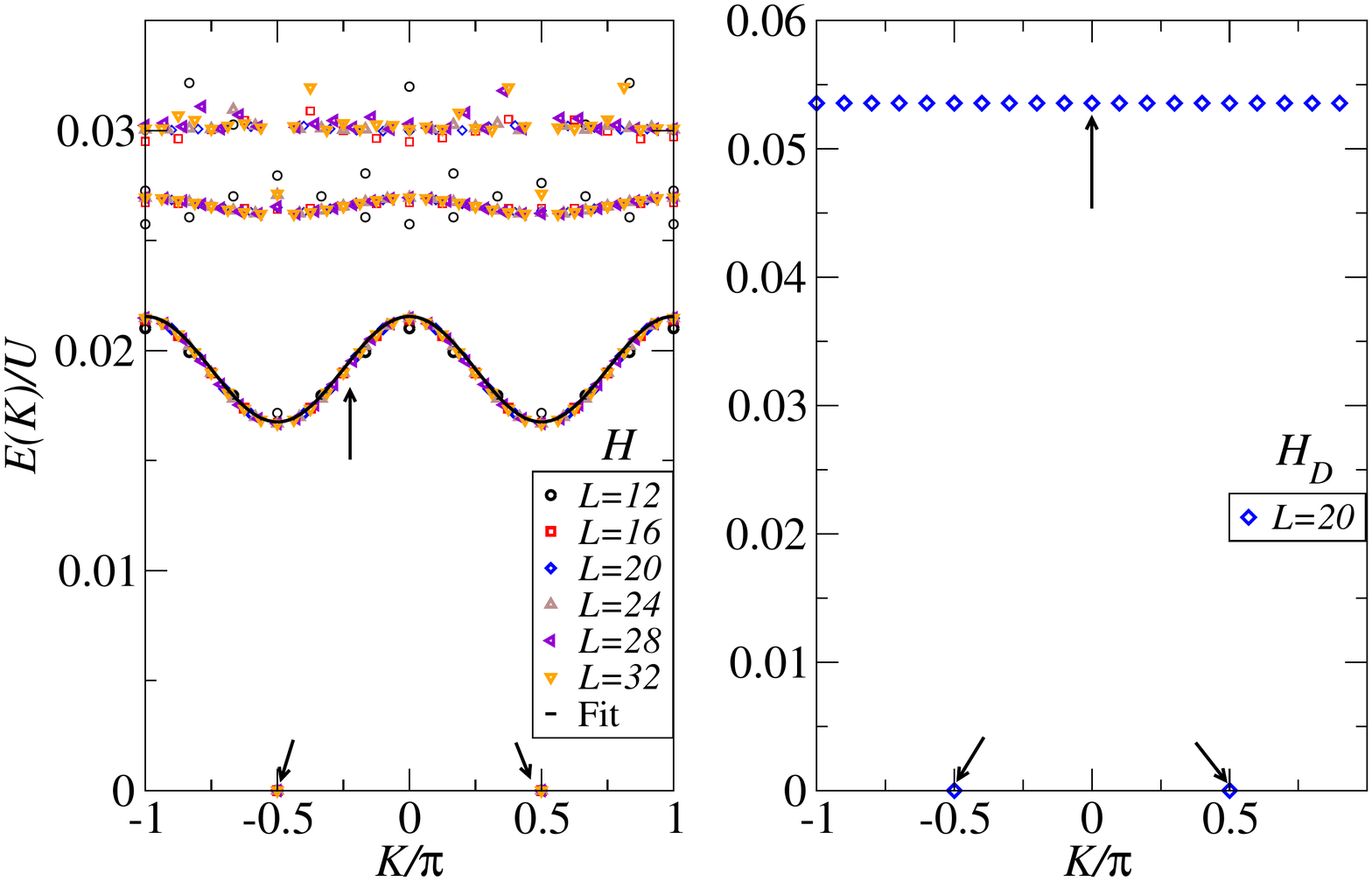}
\vspace{-1.73in}

\includegraphics[width=1.2in]{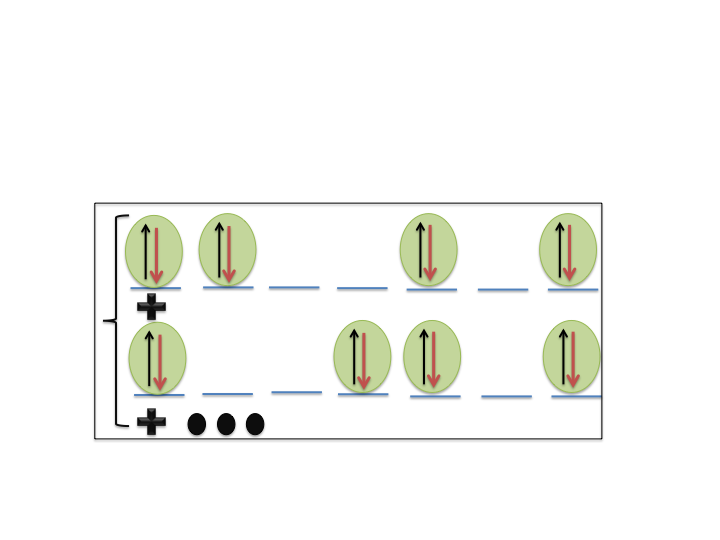}
\hspace{1.50in}
\vspace{-0.31in}

\includegraphics[width=1.25in]{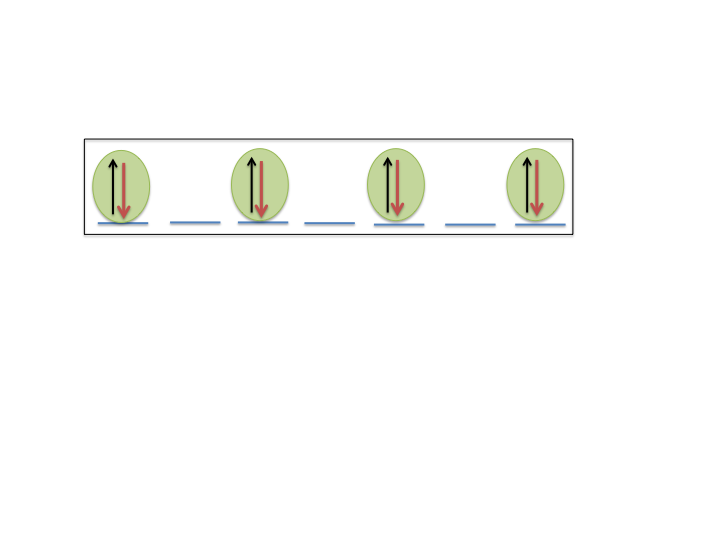}
\hspace{1.47in}

\vspace{-0.96in}
\includegraphics[width=1.25in]{wigner_crystal.png}
\hspace{-2.15in}

\vspace{-2.06in}
\includegraphics[width=1.4in]{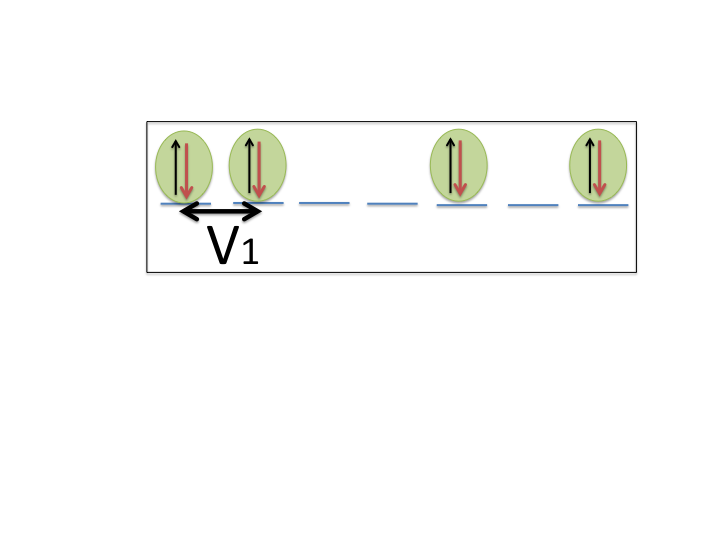}
\hspace{-1.94in}

\vspace{0.6in}

\end{center}

\caption{(Color online.) (Left Panel) The many-body energy dispersion $E(K)$ (scattered symbols) for the flatband-projected Hamiltonian $H$ on various lattice sizes, $L$. The solid curve is the fit of $E(K)$ with the effective extended Hubbard model [Eq.~(\ref{Hexthub})] that highlights a dispersive collective mode. 
(Right Panel) The energy dispersion of a similar classical model [Eq.~(\ref{DHexthub}] with $V_1/U=0.0529$, showing no dispersive collective modes.  The four schematics denote representative ground and excited state configurations of spinor $\chi$ particles (encircled arrows) in real space.  
}
\label{fitband}
\end{figure}
%%%%%%%%%%%%% %%%%%%%%%%

We numerically diagonalize Eq.~(\ref{Hflatband}) using the Lanczos algorithm. Translational symmetry allows us to work within a fixed total momentum sector, $K$.   The left panel of Fig.~\ref{fitband} shows the four lowest total energies, $E(K)$, as a function of the total momentum for several system sizes.  We find data collapse for $L\ge 16$.  Our numerics therefore apply to the thermodynamic limit.  

The ground state of $H$ is a spinor Wigner crystal shown schematically in the left panel of Fig.~\ref{fitband}, set to 
$E(\pm\pi/2)=0$.  Two Wigner crystals (both with particles at every other site) are defined in momentum space by a linear combination of wave functions at $K=\pm\pi/2$. We verify the crystalline nature of the ground state by breaking the degeneracy with a small, staggered chemical potential, $\mu\sum_i(-1)^i\tilde{n}_i$, added to Eq.~(\ref{Hflatband}).  In the $\mu\rightarrow 0\pm$ limit, the density shows that the system spontaneously picks one of the two degenerate Wigner crystal ground states \cite{supplementarymaterial}.  We have also calculated the charge
structure factor $S(k)=L^{-2}\sum_{i,j}e^{ik(r_i-r_j)}\langle \tilde{n}_i\tilde{n}_j \rangle$.  We find that $S(k)$ has well-defined peaks at $k=\pi$ , indicating Wigner crystals.

We, for comparison, numerically solve a diagonal (classical) Hamiltonian known \cite{hubbard:1978} to yield Wigner crystals:
\begin{equation}
H_D=V_1\sum_i\tilde{n}_i\tilde{n}_{i+1}.
\label{DHexthub}
\end{equation}
The right panel of Fig.~\ref{fitband} shows the many-body energy spectrum. The ground states of $H_D$ coincide with those of Eq.~(\ref{Hflatband}), i.e., at $K=\pm\pi/2$, further showing that the ground states of Eq.~(\ref{Hflatband}) are classical Wigner crystals of spinor $\chi$ particles.  The first excited state of $H_D$, however, is non-dispersive and lies at an energy $V_{1}$.  This is the energy cost of moving one particle in the Wigner crystal to a NN site (see the schematic of this classical excitation in Fig.~\ref{fitband}, right panel).   A comparison of the left and right panels shows that while the ground states are essentially the same, the excited states of  Eq.~(\ref{Hflatband}) are fundamentally different from those of Eq.~(\ref{DHexthub}).  

The excited states of Eq.~(\ref{Hflatband}), the left panel of Fig.~\ref{fitband}, exhibit a gap $\sim 0.016 U$ above the ground state.  The conditional hopping terms cause the otherwise degenerate excited band to form a dispersive collective mode.  The off-diagonal conditional hopping terms superpose the classical configurations of $\chi$ particles. (see the schematic, left panel of Fig.~\ref{fitband}).  To better understand the nature of the excited states, we construct an effective model.

%%%%%%%%% TABLE %%%%%%%%%%%%%%% 
\begin{table}[t]
  \centering
  \begin{tabular}{|c|c|c|c|c|}
    \hline\hline
    $KL/2\pi $ &  $t_1/U$ & $t_2/U$ & $|\Delta E/E_{H}|$ & $\langle \Psi_{H_{\text{eff}}}|\Psi_{H}\rangle$ \\
    \hline
    0  & 0.0117 & 0.0130 &  0.035 & 0.99 \\
    \hline
    1  & 0.0119 & 0.0135 &  0.010 & 0.97 \\
    \hline
    2  & 0.0119 & 0.0135  &  0.010  & 0.88 \\
    \hline
    3  & 0.0119 & 0.0135 & 0.015  & 0.74 \\
    \hline
    4  & 0.0119 & 0.0135 & 0.005 & 0.53 \\ 
    \hline
    5  & 0.0119 & 0.0135 & 0          & 0.99 \\
  \hline\hline
  \end{tabular}
\caption{Fitting parameters ($t_{1}$ and $t_{2}$), the resulting energy differences ($\Delta E=E_{H_{\text{eff}}}-E_{H}$), and the wave-function overlaps between $H_{\text{eff}}$ and $H$ for an $L=20$ system for the ground state at each total momentum sector, $K$, with $\mu/U=10^{-5}$. Here the small energy differences and high wave-function overlaps indicate the quality of the effective model in capturing the essential physics of the original model.} \label{bestfitparameters}
\end{table}
%%%%%%%%%%%%%%%%%%%%%%%% 

\noindent
{\it Effective Luttinger-liquid theory.} -- We construct an effective model of Eq.~(\ref{Hflatband}) by adding hopping terms to Eq.~(\ref{DHexthub}).  The effective hopping terms are emergent because they represent kinetics not present in the original model [Eq.~(\ref{Hflatband})].  We verify the accuracy of the effective model by comparing energetics and by taking wave-function overlaps.  The effective model is then studied using Luttinger-liquid theory on the emergent degrees of freedom.

\begin{figure}[t]
\vspace{-0.3in}
\includegraphics[width=3.7in]{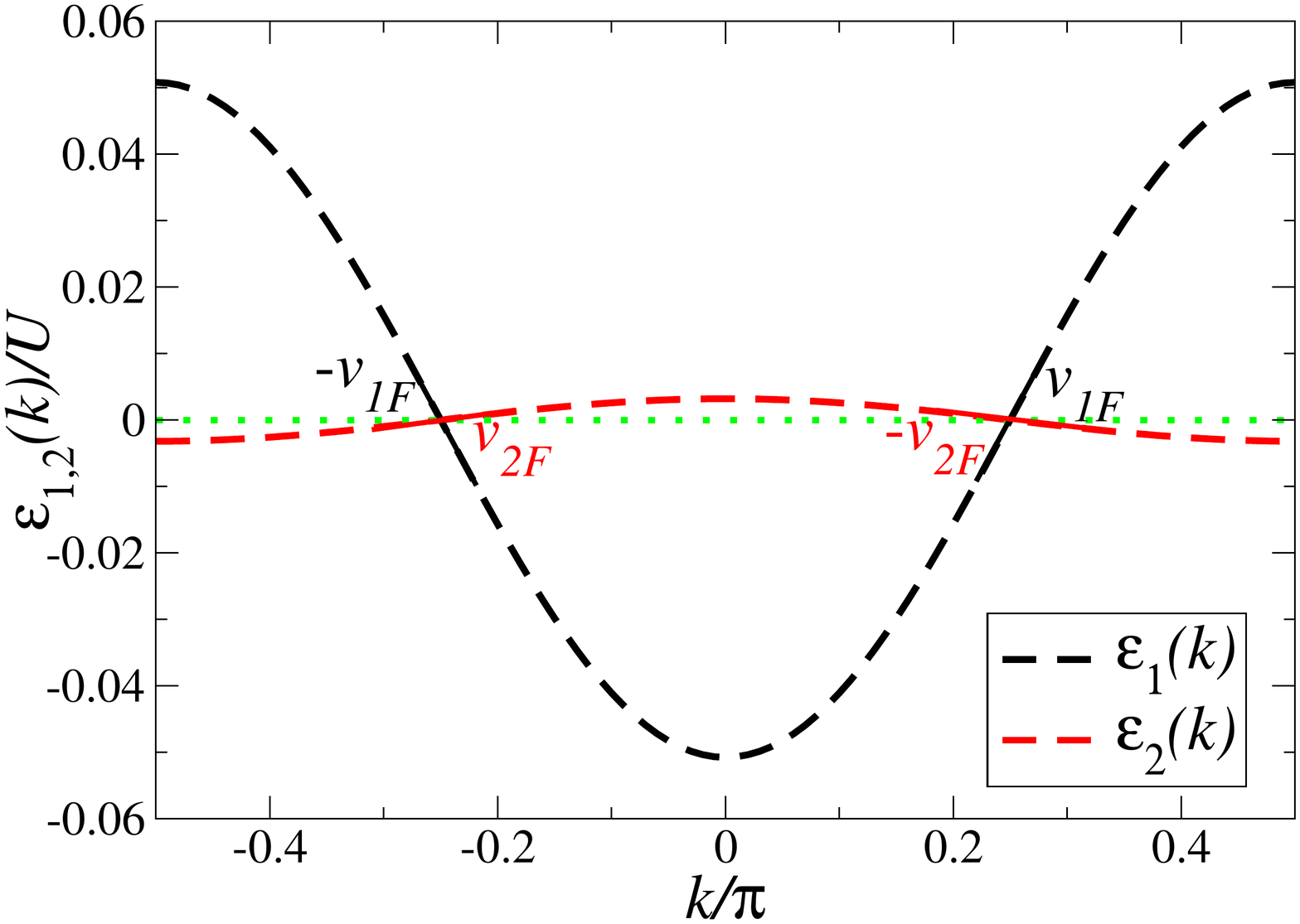}
\vspace{-2.6in}

\includegraphics[width=2.0in]{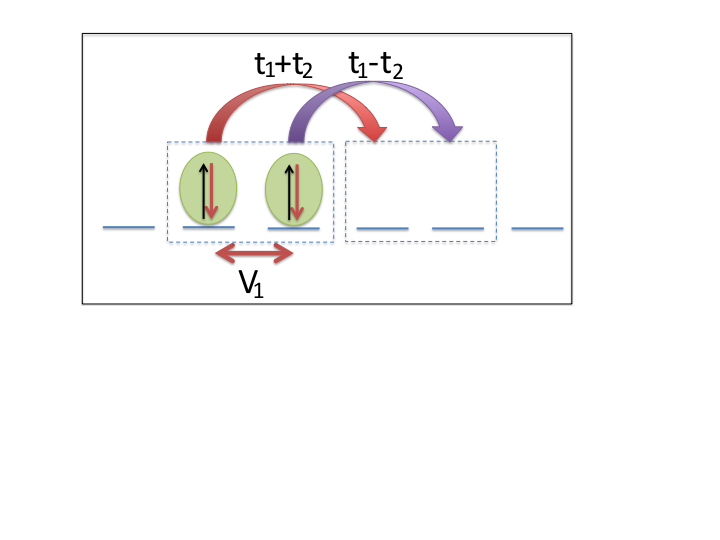}
\hspace{-0.6in}
\vspace{0.7in}

\caption{(Color online.) (Main Panel) Emergent single-particle energy dispersion, Eq.~(\ref{singleparticlenergy}).  The dashed and dotted lines cross at Fermi points, $k_{F}=\pm \pi/4$.  Linearization is shown as solid straight lines.   Differing slopes indicate asymmetric bands, i.e., $v_{1F}\neq v_{2F}$.  Inset: Schematic of hopping terms used in Eq.~(\ref{Hexthub}).}
\label{eklinearlization}
\end{figure}

We capture the effects of conditional hopping with ordinary single-particle hopping terms in an effective extended Hubbard model:
\begin{equation}
H_{\text{eff}}=-\sum_i[t_1+t_2(-1)^i](\chi_i^{\dagger}\chi_{i+2}^{\vphantom{\dagger}}+H.c.)+H_{D},
\label{Hexthub}
\end{equation}
where $t_1$ and $t_2$ are fitting parameters quantifying emergent NNN hopping.  Figure ~\ref{eklinearlization} illustrates $t_1$ and $t_2$ in real space.  Note that $t_{1}$ and $t_{2}$ scale with $U$ because we added these parameters to capture the properties of excited states generated entirely by interactions in the original hopping-free model, Eq.~(\ref{Hflatband}).     

We vary $t_1$ and $t_2$ and numerically solve Eq.~(\ref{Hexthub}) to get the best fit of $E(K)$ while maximizing overlap of the corresponding wave functions.  Table \ref{bestfitparameters} shows representative ($L=20$) fits for the lowest eigenstates. The energy differences between $H_{\text{eff}}$ and $H$ are all within $5\%$, and the 
wave-function overlaps for the lowest states are all above $50\%$ with almost $100\%$ overlaps at $K=0$ and $\pm\pi$.  The overlap between the ground states ($K=\pm\pi/2$) is above $99.9\%$. We plot the first excited state of Eq.~(\ref{Hexthub}) from the best fit parameters as the black curve in the left panel of Fig.~\ref{fitband} for comparison.  The overlap and energetic comparison show that Eq.~(\ref{Hexthub}) captures the essential properties of Eq.~(\ref{Hflatband}) at low energies.   We can therefore use Eq.~(\ref{Hexthub}) as an effective theory to make predictions for experiments. 
 
We now show that Eq.~(\ref{Hexthub}) exhibits excitations with fractionalized charge quantified by Luttinger-liquid theory.  We first diagonalize the hopping terms in Eq.~(\ref{Hexthub}) \cite{supplementarymaterial}.   The emergent ``single-particle'' energy dispersion has two energy bands $(b=1,2)$:
\begin{equation}
\varepsilon_b(k)=-2\big[t_1-(-1)^b t_2\big ]\cos(2k),
\label{singleparticlenergy}
\end{equation}
with Fermi velocities $v_{bF}\equiv \vert \partial\varepsilon_b/\partial k \vert_{k_{F}} =|4\big [t_1-(-1)^{b}t_2\big ]\sin(2k_{F})|$.  For $N/L=1/2$ each dispersion crosses the Fermi level at two Fermi points $ k_F=\pm\pi/4$ (Fig.~\ref{eklinearlization}).   Low-energy excitations near the Fermi points therefore consist of two left movers and two right movers.

%%%%%%%%%%SINGLE PARTICLE BANDS%%%%%%%%%%%%%% 

%%%%%%%%%%%%%%%%%%%%%%%% 

We bosonize Eq.~(\ref{Hexthub}) to study interaction effects.   We linearize the dispersion at the Fermi points \cite{haldane:1981, gimarchy:2003, miranda:2003}, as depicted in Fig.~\ref{eklinearlization}.  The elementary excitations near $\pm k_F$ are bosonic and, in the absence of the interacting term in Eq.~(\ref{Hexthub}), have the charge of the original flat band particles.  We include $H_{D}$ and find the normal modes of the bosonized Hamiltonian using a unitary transformation and rescaling of the bosonic fields \cite{sedlmayr:2013}. The emergent normal mode 
Luttinger parameters, i.e., velocity, $u_{l}$, and the charge fractionalization ratio, $g_{l}$, are given by \cite{supplementarymaterial}
\begin{eqnarray}
u_{l}&=&(v_{1F}v_{2F}\tilde{\lambda}_{l})^{\frac{1}{2}},\\
g_{l}&=&(\lambda_{l}\tilde{\lambda}_{l})^{-\frac{1}{2}},
\label{luttingerparameters}
\end{eqnarray}
where $l=1,2$ denotes the two normal modes, $\lambda_1\equiv v_{1F}/v_{2F}$, $\lambda_2\equiv\lambda_1^{-1}$, and 
$\tilde{\lambda}_{l}\equiv \Big [\lambda_1+\lambda_2-(-1)^l\sqrt{(\lambda_1-\lambda_2)^2+4V_1^2/(\pi^2 v_{1F}v_{2F}})\Big ]/2$.
For $V_1>0$, we have $g_1<1$ indicating that the charge has fractionalized for this normal mode.  To see this we write the effective charge, $q^{*}$, in terms of the original charge, $q$, as $q^{*}=g q$ where $q^{*}$ can be inferred from particle number conductance \cite{fisher:1997}.  $g_1$ found here can be continuously tuned below unity.  This should be contrasted with fractionally charged excitations in the FQH regime, where the fractions are only rational \cite{laughlin:1983,goldman:1995,dassarma:1997}. 

The Luttinger-liquid analysis therefore shows that low-energy collective modes of Eq.~(\ref{Hexthub}) can be thought of as fractionalized quasiparticles moving along a spinor Wigner crystal.  This result, while known in standard Luttinger-liquid theories \cite{haldane:1981,fisher:1997,miranda:2003,gimarchy:2003}, is surprising here since the single-particle eigenstates of the physical atoms are inert (flatband) particles that derive emergent kinetics from interactions.  The close connection between Eqs.~(\ref{Hexthub}) and  (\ref{Hflatband}) also indicates that these modes should be experimentally observable.

\noindent
\emph{Experimental requirements and observables.} -- Low temperatures and low atomic losses are, in general, difficult requirements for proposals to engineer strongly correlated quantum states with atomic gases.  Most proposals require maximizing $U$ by tuning a Feshbach resonance to enter strongly correlated regimes.  However, Feshbach resonances contribute to unwanted heating and losses \cite{stenger:1999}, particularly in SOC atomic gases \cite{wei:2013,williams:2013}.  The flat band regime studied here circumvents the need for strong $U$ (and therefore a Feshbach resonance) because the system is automatically strongly correlated in the absence of kinetic energy.  

We can estimate realistic parameters to show that the flatband regime is attainable. $^{40}$K is one of the best candidates for strong SOC with low losses \cite{wei:2013,williams:2013}. For $^{40}$K in a 1D optical lattice with $s=10$, $k_R=k_L/2$, $\Omega=0.05 E_{R}$, and a perpendicular confinement of lattice depth $60 E_{R}$, we find  $\Delta_{SO} \approx 0.10 E_{R}$, $V_{1}\approx 0.014 E_{R}$ and $W \approx 0.013 E_{R}$, where $W=4t$ is the bandwidth.  This shows that even bare $s$-wave scattering implies a strongly interacting flatband problem with $\Delta_{SO} \gg V_{1} \gtrsim W$.  Note that the last inequality is a very stringent flatband requirement.  An accurate (but weaker) requirement assumes the many-body energy gap $\approx V_{1}/3$ (left panel, Fig.~\ref{fitband}) to be larger than the single-particle hopping $V_{1}/3 \gtrsim W/4$.  This implies that partial filling of the lowest band allows us to treat Eq.~(\ref{spinorbitH}) as an irrelevant constant for realistic system parameters.  

Parabolic confinement will compete with the many-body energy gap to diminish the size of the Wigner crystal near the trap center.  The central crystal will give way to edge states when the parabolic trapping potential energy reaches the gap, i.e., $V_{1}/3\approx m\omega_{\text{tr}}^2x_{\text{max}}^2/2$.  The crystal will then be as large as $\approx2\sqrt{2V_1/3m\omega_{\text{tr}}^2}$ sites.  Trapping potentials therefore place a lower bound on the size of the energy gap (and therefore $U$).  For $^{40}$K we find that even the bare $s$-wave scattering length allows significant crystal sizes, $\sim 86-150$ sites for realistic trapping strengths, $\omega_{\text{tr}}=40- 70$Hz.  Larger interaction strengths will increase the size of the crystal.  

Observations of the states proposed here are in principle possible with currently available methods.  The spinor Wigner crystal state manifests as a peak in the static structure factor of the original fermions, observable with demonstrated probes:  noise correlations \cite{altman:2004} or atomic matter wave scattering \cite{gadway:2012}.  Luttinger-liquid parameters have also been observed by interfering Bose-Einstein condensates \cite{hofferberth:2008}.  Detecting fractionalized charge is more challenging.  In the current context, it could be measured by, e.g., detection of partial backscattering from an impurity \cite{fisher:1997,daley:2008}, optical methods \cite{ruostekoski:2002,javanainen:2003}, or charge pumping \cite{qian:2011,wang:2013}.

\noindent
{\it Summary.} -- We predict a set of intriguing collective states of matter in experiments with atomic Fermi gases confined to 1D optical lattices and in the presence of synthetic SOC.   We constructed and studied a model where the atomic interactions operate in a flatband.   We found that the single-particle basis states are delocalized spinors.  Our analysis predicts that flatband spinor particles have surprising properties generated by on-site interactions among the original atoms: NN interactions and effective NNN hopping.  The many-body ground state was found to be a Wigner crystal of spinors.  We find that an effective Luttinger-liquid theory parametrizes emergent kinetics and fractionalized charge [Eq.~(\ref{luttingerparameters})] in the low-energy collective modes, in direct analogy to the mechanism of emergence found in the FQH regime.    

We acknowledge helpful comments from I. Spielman and support from the ARO (W911NF-12-1-0335,W911NF-12-1-0334), AFOSR (FA9550-11-1-0313), and DARPA-YFA.

\section{Supplementary Material for ``Emergent Kinetics and Fractionalized Charge in 1D Spin-Orbit Coupled Flatband Optical Lattices''}

\renewcommand{\thepage}{S\arabic{page}}
\renewcommand{\thesection}{S\arabic{section}}
\renewcommand{\thetable}{S\arabic{table}}
\renewcommand{\thefigure}{S\arabic{figure}}
\renewcommand{\theequation}{S\arabic{equation}}

\subsection{Flat band Projected Model}

In this section we derive the flat band projected Hamiltonian, Eq.~(2), by adding interactions to Eq.~(1).
We first use the tight-binding limit to show that flat bands arise from spin-orbit coupled fermions in optical lattices in the absence of interactions.  Using the tight-binding limit we transform the interaction to an on-site Hubbard interaction.  We then project the interaction into the flat band to derive Eq.~(2).
We conclude by showing the limits in which the flat band approximation is accurate.

We start by noting that Eq.~(1) yields Bloch bands that can be accurately fit with a tight-binding model (for $s>2$):
\begin{equation}
H_{TB}=-2t\sum_{k}C_k^{\dagger}\cos(k+k_R\sigma_z)C_k^{\phantom{\dagger}}+\Omega \sum_{k}C_k^{\dagger}\sigma_x C_k^{\phantom{\dagger}},
\label{tightbindH}
\end{equation}
where the NN hopping $t$ is tuned to fit the exact band width of the first Bloch band of Eq.~(1) with $k_{R}=0$ and $\Omega=0$.  $t$ is therefore a single-particle hopping derived entirely from optical lattice parameters.  Here $C_k^{\dagger}=(c_{k\uparrow}^{\dagger}, c_{k\downarrow}^{\dagger})$ creates a fermion spinor.

We use Eq.~(\ref{tightbindH}) to derive the flat band projected Hamiltonian.  Eq.~(\ref{tightbindH}) can be solved exactly.  The two lowest eigenvalues are given by:
\begin{equation}
\omega_{1,2}(k)=-2t\cos k\cos k_R\pm\sqrt{4t^2\sin^2 k\sin^2 k_R+\Omega^2}.
\label{eigenvalues}
\end{equation}
 Fig.~1 plots the lowest bands.  We have tuned the spin-orbit coupling strength
$k_R$ and the Zeeman field strength $\Omega$ to generate flat bands.  For both lattice depths we chose the same band flatness ratio \cite{zhang:2013}, $F\equiv \Delta_{\text{SO}}/W\approx 8$, where $W=4t$ is the band width and $\Delta_{SO}$ is the separation between the two lowest bands.

The unitary matrix that diagonalizes Eq.~(\ref{tightbindH}) is:
\begin{equation}
M=\begin{bmatrix}
\cos(\alpha_k) & -\sin(\alpha_k) \\
\sin(\alpha_k)  &  \cos(\alpha_k) \\
\end{bmatrix},
\end{equation}
where
\begin{eqnarray}
\sin(\alpha_k)&=&\frac{\omega_1(k)-h(k)}{\sqrt{\Omega^2+(\omega_1(k)-h(k))^2}},\\
\cos(\alpha_k)&=&\frac{\Omega}{\sqrt{\Omega^2+(\omega_1(k)-h(k))^2}}.
\end{eqnarray}
and $h(k)\equiv-2t\cos(k+k_R)$.  The eigenstates of Eq.~(\ref{tightbindH}) can then be written in terms of the original fermi operators:
\begin{equation}
\begin{bmatrix}
\zeta_k \\
\chi_k\\
\end{bmatrix}=M^{\dagger}
\begin{bmatrix}
c_{k,\uparrow}\\
c_{k,\downarrow}\\
\end{bmatrix},
\label{Utransform}
\end{equation}
where $\chi_k$ and $\zeta_k$ denote operators for lower and upper Bloch bands, respectively (Fig.~1).

We can now use the unitary transformation to study the addition of interactions to the non-interacting model within the lowest flat band.
Occupancy of just the lowest flat band implies that Eq.~(\ref{tightbindH}) acts as an irrelevant constant, to a first approximation.  The relevant term in the model derives from interactions between atoms.  In the tight-binding limit, $s$-wave contact interactions thus yield an on-site projected Hubbard interaction:
\begin{eqnarray}
H=\text{const.}+U\sum_i \mathcal{P} c_{i\uparrow}^{\dagger}c_{i\downarrow}^{\dagger}c^{\vphantom{\dagger}}_{i\downarrow}c^{\vphantom{\dagger}}_{i\uparrow}\mathcal{P}, \label{Hinteraction}
\end{eqnarray}
where $U$ is the on-site repulsion, $\mathcal{P}$ projects particles into the lowest flat band, $c_{i\uparrow}^{\dagger}$ creates a fermion at lattice site $i$ in the $\uparrow$ spin state, and the constant is the zero-point energy of the flat band.  By using Eq.~(\ref{Utransform}) in combination with a Fourier transform, Eq.~(\ref{Hinteraction}) becomes:
\begin{eqnarray}
H=U\sum_{\{k\}}f_{\{k\}}\chi_{k_1}^{\dagger}
\chi_{k_2}^{\dagger}\chi_{k_3}^{\phantom{\dagger}}\chi_{k_4}^{\phantom{\dagger}},
\label{Hflatband}
\end{eqnarray}
where $f$ is defined in the main text.  This shows that Eq.~(2), follows from the on-site interactions operating in a flat spin-orbit band.

We now argue that all fermions occupy the lowest band in realistic parameter regimes.  This is a valid assumption provided the system only partially fills the lowest band and $\Delta_{SO}$ is greater than inter-band interaction matrix elements.   The $s$-wave interaction can be tuned far from a Feshbach resonance to ensure that $\Delta_{SO}$ is larger than characteristic interaction energies.  This maintains the flat band condition provided $U\gg W$.  We note that, in a perfectly flat band $W\rightarrow0$, the problem remains strongly correlated even for small $U$ because the interaction becomes the only term in the Hamiltonian.   The main text shows that $\Delta_{SO} \gg V_{1} \gtrsim W$ is satisfied for $^{40}$K.

\subsection{Detecting Spontaneous Symmetry Breaking}
In this section we show that application of a small staggered chemical potential spontaneously selects one of the degenerate Wigner crystal states.
The lowest energies from our exact diagonalization study of $H$ [Eq.~(2)] are 2-fold degenerate at total momentum $K=\pm\pi/2$.  Each state is uniform in the absence of symmetry breaking.  If the ground state is truly uniform, a small perturbation (much smaller than the gap) should leave the ground state density intact.  If, however, the ground state prefers to spontaneously select an inhomogeneous configuration, a small non-uniform perturbation should drive the system into one of the crystal configurations.  In our case (Eqs.~2 or 3) a logical choice for degenerate crystal configurations has one particle at every other site in real space.  The crystals are defined as linear combinations of the two lowest energy states at $K=\pm\pi/2$.

To identify crystalline order we perturb the ground state with a small staggered chemical potential term $\mu\sum_i(-1)^i\tilde{n}_i$ to check if the ground
state tends to spontaneously choose one of the two degenerate Wigner crystal states. (This is similar to methods employed in the numerical determination of staggered magnetization in the antiferromagnetic Ising model, where a small staggered magnetic field is also applied to pick one of the two staggered magnetization directions.)

Care needs to be taken in the limiting procedure. For a fixed small positive $\mu$, one needs to extrapolate first the lattice size, i.e., $L\rightarrow\infty$. The ground state selected as $\mu\rightarrow 0+$ then denotes one of the two broken-symmetry Wigner crystal states. The other Wigner
crystal state can be detected in the limit $\mu\rightarrow 0-$.  This shows that spontaneous breaking of the discrete sublattice symmetry inherent in Wigner crystals can be detected with the application of a small staggered chemical potential.

\subsection{Derivation of Emergent Luttinger Parameters}

In this section we prove the formulas for the emergent Luttinger parameters $u_{l}$ and $g_{l}$ in the main text, Eqs.~(6) and (7).  To do this we apply Luttinger liquid theory to the effective model in the main text.  We first bosonize the non-interacting part of the model and then the interacting part.

The effective model [Eq.~(4)] is:
\begin{equation}
H_{\text{eff}}=H_{0}+H_{D}
\label{Hexthubeq}
\end{equation}
where the non-interacting part can be rewritten in terms of two-component vectors in $k$-space:
\begin{equation}
H_{0}=-2(t_1+t_2)\sum_k^{'}\cos(2k)\xi^{\dagger}_k(I+\sigma_x)\xi_k^{\vphantom{\dagger}}.
\label{noninteractionH}
\end{equation}
Here the prime indicates summation over $[-\pi/2, \pi/2)$, $I$ is the identity matrix, $\sigma_x$ is the $x$-component of the Pauli matrix, and $\xi^{\dagger}_k=(\chi_{k}^{\dagger},\chi^{\dagger}_{k-\pi})$.

We diagonalize $H_{0}$ with a unitary transformation, defined on a restricted momentum range $[-\pi/2, \pi/2)$:
\begin{eqnarray}
\chi_k&=&(\chi_{1k}-\chi_{2k})/\sqrt{2},\nonumber\\
\chi_{k-\pi}&=&(\chi_{1k}+\chi_{2k})/\sqrt{2}.
\label{chi_transform}
\end{eqnarray}
Here $1$ and $2$ label the two bands established by the sublattice dependent hopping in the effective model.  The non-interacting Hamiltonian then becomes:
\begin{equation}
H_0=\sum_{b=1,2}\sum_k^{'}\varepsilon_{b}(k)\chi_{bk}^{\dagger}\chi_{bk}^{\vphantom{\dagger}},
\label{Hochi}
\end{equation}
where $\varepsilon_{b}$ is defined in the main text [Eq.~(5)].

To begin the bosonization process, we pass to the continuum limit and expand the field operators into left and right moving fermion fields around the two Fermi points:
\begin{eqnarray}
\chi_{1i}&\rightarrow&e^{ik_Fx}\psi_{1R}+e^{-ik_Fx}\psi_{1L},\\
\chi_{2i}&\rightarrow&e^{-ik_Fx}\psi_{2R}+e^{ik_Fx}\psi_{2L},
\end{eqnarray}
where we have taken into account the opposite slopes in linearizing the effective single-particle energy dispersions of two bands (see Fig.~3).  We can use the left/right (L/R) decomposition to bosonize Eq.~(\ref{Hochi}) using the usual bosonization methods.  We define four bosonic fields, $\phi_{br}(x)$ with $b=1,2$ and $r=R,L$, such that:
\begin{eqnarray}
\psi_{br}^{\vphantom{\dagger}}(x)&\sim&\frac{1}{\sqrt{2\pi\epsilon}}e^{-i\sqrt{2\pi}\phi_{br}(x)},\\
:\psi_{br}^{\dagger}(x)\psi_{br}^{\vphantom{\dagger}}(x):&=&\mp\frac{1}{\sqrt{2\pi}}\partial_x\phi_{br}(x),
\end{eqnarray}
where ``$:\phantom{} :$'' indicates normal ordering, $\epsilon\rightarrow 0^{+}$, and $+(-)$ corresponds to $r=L (r=R)$.

With these transformations, Eq.~(\ref{noninteractionH}) becomes:
\begin{equation}
H_0\approx\sum_{b,r}\frac{v_{bF}}{2}\int_{-L/2}^{L/2}dx(\partial_x\phi_{br})^2.
\end{equation}
This form for $H_{0}$ explicitly reveals the four-component nature of the excitations near the Fermi points: two bands ($b=1,2$) and two directions of motion ($r=L,R$).

We now apply the bosonization procedure to the interacting term in Eq.~(\ref{Hexthubeq}), $H_{D}$.
We first Fourier transform the interaction, apply the unitary transformation using Eq.~(\ref{chi_transform}), and we finally substitute the Fourier transform for $\chi_{i}^{\dagger}$ into $H_{D}$.  This
gives rise to 256 terms.   Most of these terms cancel to yield:
\begin{eqnarray}
H_D&=&V_1\sum_i^{\rm{even}}\chi_{1,i}^{\dagger}\chi_{1,i}^{\vphantom{\dagger}}\chi_{2,i+1}^{\dagger}\chi_{2,i+1}^{\vphantom{\dagger}}\nonumber\\
&+&V_1\sum_i^{\rm{odd}}\chi_{2,i}^{\dagger}\chi_{2,i}^{\vphantom{\dagger}}\chi_{1,i+1}^{\dagger}\chi_{1,i+1}^{\vphantom{\dagger}}, 
\end{eqnarray}
where we have made use of the fact that even and odd sites correspond to bands 1 and 2, respectively.

We are now able to bosonize the interaction term using the same transformations as those used above.  We first note that, since we are working at half filling ($k_{F}=\pm \pi/4$), Umklapp terms will vanish. We further define two conjugate fields for the two bands:
\begin{eqnarray}
\phi_b&=&(\phi_{bL}-\phi_{bR})/\sqrt{2},\nonumber\\
\theta_b&=&(\phi_{bL}+\phi_{bR})/\sqrt{2}.
\end{eqnarray}
The total bosonized Hamiltonian can then be written in terms of the conjugate fields:
\begin{eqnarray}
H_{\text{eff}} \approx \frac{\sqrt{v_{1F}v_{2F}}}{2}\int_{-L/2}^{L/2}dx \left [ \Phi^T M_{\phi}\Phi + \Theta^T M_{\theta}\Theta\right ]
\end{eqnarray}
where
$\Phi^T\equiv(\partial_x\phi_1, \partial_x\phi_2)$ and $\Theta^T\equiv(\partial_x\theta_1, \partial_x\theta_2)$. The two matrices $M_{\phi}$ and $M_{\theta}$ are given by:
\begin{eqnarray}
M_{\phi}&=&
\begin{bmatrix}
\sqrt{\frac{v_{1F}}{v_{2F}}} & \frac{V_1}{\pi \sqrt{v_{1F}v_{2F}}} \\
\frac{V_1}{\pi \sqrt{v_{1F}v_{2F}}}  & \sqrt{\frac{v_{2F}}{v_{1F}}} \\
\end{bmatrix},\nonumber\\
M_{\theta}&=&
\begin{bmatrix}
\sqrt{\frac{v_{1F}}{v_{2F}}} & 0 \\
0  & \sqrt{\frac{v_{2F}}{v_{1F}}} \\
\end{bmatrix}.
\end{eqnarray}

To find the emergent Luttinger parameters we diagonalize the matrices defining the above Hamiltonian with the following transformation \cite{sedlmayr:2013}:
\begin{eqnarray}
\Phi=T_{\phi}\tilde{\Phi} \text{  and  } 
\Theta=T_{\theta}\tilde{\Theta},
\label{T_transform}
\end{eqnarray}
where the transformation matrices satisfy $[T_{\phi}]^T=[T_{\theta}]^{-1}$.  This guarantees that the new conjugate fields that hybridize the bands into normal modes, $\tilde{\Phi}$ and $\tilde{\Theta}$, satisfy the canonical commutation relations.

It can be checked that the above condition is fulfilled by the following choice of transformation matrices:
\begin{eqnarray}
T_{\phi}=\Lambda^{-1}Q\tilde{\Lambda}^{-1} \text{  and  } 
T_{\theta}=\Lambda Q\tilde{\Lambda},
\end{eqnarray}
where the unitary matrix $Q$ diagonalizes the rescaled matrix $M_{\phi}^{'}=\Lambda^{-1}M_{\phi}\Lambda^{-1}$, i.e.,
\begin{equation}
Q^T M_{\phi}^{'}Q=
\begin{bmatrix}
\tilde{\lambda}_1  & 0 \\
0  & \tilde{\lambda}_2 \\
\end{bmatrix},
\end{equation}
and the diagonal rescaling matrices $\Lambda$ and $\tilde{\Lambda}$ are given by:
\begin{eqnarray}
\Lambda&=&
\begin{bmatrix}
(\lambda_1)^{-\frac{1}{4}} & 0 \\
0  & (\lambda_2)^{-\frac{1}{4}} \\
\end{bmatrix},\\
\tilde{\Lambda}&=&
\begin{bmatrix}
(\tilde{\lambda}_1)^{\frac{1}{4}}  & 0 \\
0  & (\tilde{\lambda}_2)^{\frac{1}{4}}
\end{bmatrix}.
\end{eqnarray}
Here $\lambda_{1,2}$ and $\tilde{\lambda}_{1,2}$ are defined in the main text.

The total Hamiltonian defining the effective Luttinger liquid theory can then be written in terms of the diagonalized Hamiltonian for each of the normal modes $(l=1,2)$:
\begin{equation}
H_{\text{eff}}\approx \sum_{l}\frac{u_l}{2}\int_{-L/2}^{L/2}dx\left [g_l(\partial_x\tilde{\theta}_l)^2+\frac{1}{g_l}(\partial_x\tilde{\phi}_l)^2 \right ],
\end{equation}
where the emergent Luttinger parameters $u_l$ and $g_l$ are given in the main text, Eqs.~(6) and (7).

\end{document}